\documentstyle[aps,preprint]{revtex}

\begin{document}
\draft
\preprint{manuscript}
\title{Directional dependence of spin currents induced by 
Aharonov-Casher phase}
\author{Taeseung Choi$^{a}$, Chang-Mo Ryu$^{a}$, and 
A. M. Jayannavar$^{b}$}
\address{$^{a}$ Department of Physics,
Pohang University of Science and Technology,
         Pohang 790-784, South Korea \\
$^{b}$ Institute of Physics, Sachivalaya Marg,
         Bhubaneswar-751005, India
         }
\date{\today}
\maketitle

\begin{abstract}
We have calculated the persistent spin current of an open ring 
induced by the Aharonov-Casher phase.
For unpolarized electrons there exist no persistent charge currents,
but persistent spin currents.
We show that, in general, the magnitude of the persistent spin
current in a ring depends on the direction of the direct current flow
from one reservoir to another.
The persistent spin current is modulated by the cosine function
of the spin precession angle.
The nonadiabatic Aharonov-Casher phase gives anomalous behaviors.
The Aharonov-Anandan phase is determined by the solid angle of spin 
precession.
When the nonadiabatic Aharonov-Anandan phase approaches a constant 
value with the increase of the electric field, the periodic behavior of
the spin persistent current occurs in an adiabatic limit. 
In this limit the periodic behavior of the persistent spin current  
could be understood by the effective spin-dependent Aharonov-Bohm
flux.

\end{abstract}


\narrowtext
The electric and magnetic properties of mesoscopic systems have recently
received much attention in the light of several experimental 
observations \cite{Kram,Alt,Been,Wash}.
Mesoscopic physics deals with the structure made of metallic or 
semiconducting material on a nanometer scale.
The length scale associated with the dimensions in these systems are
much smaller than the inelastic mean free path or phase breaking length.
In this regime, an electron maintains phase coherence across
the entire sample.
In general, a system with a large degree of freedom is called mesoscopic
if the length up to which the wave function retains phase coherence 
exceeding the size of the system.
The main characteristics of mesoscopic systems is the quantum coherence.
These systems, which are now accessible experimentally, provide an ideal
test ground for the quantum mechanical models beyond the atomic realm.
These systems have revealed, several interesting and previously 
unexpected quantum effects at low temperatures 
\cite{Kram,Wash,Beaum,Fuk}, which are associated with the quantum 
interference of electron waves, quantization of energy levels, 
and discreteness of electron charge.
Persistent currents in mesoscopic normal metal rings are purely 
mesoscopic effects in the sense that they are strongly suppressed 
when the ring size exceeds the characteristic dephasing length of 
the electrons or the inelastic mean free path \cite{Chand,Mail}.
Studies have been extended to include multichannel rings, spin-orbit
coupling, disorder, electron-electron interaction effects, etc.
\cite{Kram,Cheu,Entin}.

Theoretical treatments up to date have been mostly concentrated on
isolated rings. Persistent current occur not only in isolated rings but
also in the rings connected via leads to electron reservoirs, namely
open systems \cite{Butt1,Butt2,Jay1}.
In a recent experiment Maily et al. have measured the persistent 
currents in both closed and open rings \cite{Mail}.
Recently Jayannavar et al. noted the several novel effects related to
persistent currents can arise in open systems, which have no analogue in
closed or isolated systems \cite{Jay2,Jay3,Par,Jay4}.
Especially the directional dependence of persistent current in open 
system can be useful for separating the persistent current from noises.

In 1984, Aharonov and Casher (AC) \cite{b4} noticed the possibility of 
the dual effect of the AB phase and discovered the AC phase for a 
neutral magnetic moment encircling a charged line.
In a fundamental generalization of Berry's idea \cite{b1}, Aharonov and
Anandan (AA) removed the adiabatic restriction and studied the geometric
phase for the nonadiabatic cyclic evolution \cite{b12}.
By removing the dynamical part, Aharonov and Anandan
defined the nonadiabatic geometric phase for the cyclic evolution
called the AA phase.
Qian and Su \cite{b13} has demonstrated the existence of
the AA phase in the AC effect.
In the adiabatic limit this AA phase becomes the spin-orbit Berry phase
introduced by Aronov and Lyanda-Geller  \cite{b14}.
Loss, Goldbart, and Balatsky discovered that Berry phase can induce
persistent spin currents \cite{Los}. And Balatsky and Altshuler noticed
spin-orbit interaction produces persistent spin and mass currents 
\cite{Bal}.
Along this line of study of the spin phase effects on the electron
transport problem, Ryu \cite{Ryu} has shown that various spin motive
forces \cite{Ster} can be described in a unified fashion
based on the Goldhabor-Anandan \cite{Gold} gauge theory for a 
low energy spin particle.
The persistent current induced by the Aharonov-Casher (AC) phase
is much smaller than the persistent current induced by the 
Aharonov-Bohm (AB) phase, so the directional dependence will be 
extremely useful for the detection of that current.
The transport behavior  induced by the AC phase is recently 
studied \cite{Qia,Ours}.

In our present treatment we consider a one-dimensional metal loop
of length $L$ coupled to two electron reservoirs as shown in Fig. 1.
In the ring there is a cylindrically symmetric electric field to produce
a spin-orbit interaction.
This spin-orbit interaction gives the AC phase with cyclic evolution.
This idealization to one-dimension corresponds experimentally to
a network of high-mobility quantum wires with narrow width such that
only the lower subband is filled. 
Our calculations are for noninteracting systems of electrons.
In such a geometry the AC effect manifests itself not only in a 
transport phenomenon but also in a persistent current. 
The left and right reservoirs are characterized by chemical potentials 
$\mu_1$ and $\mu_2$, respectively.  
We have introduced a $\delta$-function impurity of strength $V$ at a 
length $L_d (=2L)$ to the right of the metal loop 
(marked by $\times$ in Fig. 1).
The presence of the impurity breaks the spatial symmetry of the system.
We also restrict to the case of $L_1 = L_2$, to avoid the additional
contribution arising due to the difference in transport current
across upper and lower arms.
If $\mu_1 > \mu_2$ the net current flows from the left to the right
and vice versa, if $\mu_1 < \mu_2$.
The scattering of the electronic wave function occurs at the junctions
$J_1$, $J_2$ and at the impurity site $I$.
In our model we have complete spatial separation between elastic 
processes in the loop and the inelastic processes in the reservoirs.
The inelastic processes in the reservoir are essential to obtain a 
finite conductance.

When $\mu_1 > \mu_2$, the steady flux of electrons with an energy $E$
is injected from the reservoir $1$.
These electrons moving to the right are first scattered at the 
junction $J_1$ and subsequently at $J_2$ and $I$ (together with 
multiple reflections at $J_1$, $J_2$, and $I$).
The electrons emitted by the reservoir $2$ are first scattered at 
$I$ and subsequently at $J_2$ and $J_1$.
Since there is no spatial symmetry, for these two different cases the 
electron wave function (scattering states) has a different complex 
amplitude at $J_1$ and $J_2$.
The persistent current in a metallic loop is sensitive to the boundary
condition, and hence we observe that the magnitude of the persistent
current depends on the direction of the current flow.
Obviously the conductance of an entire network (calculated via the 
quantum transmission coefficient) does not depend on the direction of 
the current flow. 
This implies that there is no simple scaling relation between the
persistent currents and the conductance of the entire network.

First we consider the situation wherein the direct current flows
from the left reservoir to the right reservoir.
In the presence of cylindrically symmetric electric fields ${\bf E}$,
the one-particle Hamiltonian for non-interacting electrons is given by
\begin{equation}
H = \frac{1}{2m_e} ({\bf p} - \frac{\mu}{c} \mbox{\boldmath $\sigma$} 
\times {\bf E} )^2,
\end{equation}
where $\mbox{\boldmath $\sigma$} \times \frac{{\bf E}}{2}$ represents 
a spin-orbit coupling and $\sigma^{\alpha}$ with $\alpha =1, 2, 3$ are 
Pauli matrices.
Adopting a cylindrical coordinate system and the electric field 
${\bf E} = E( \cos \chi \hat{r} - \sin \chi \hat{z})$ we have the 
following Hamiltonian in a closed ring

\begin{equation}
H= \frac{\hbar^2}{2m_e a^2} \left(-i \partial_{\phi} - 
\frac{\mu E a}{2 \hbar c}
(\sin{\chi} \cos{\phi} \sigma_x + \sin{\chi} \sin{\phi} \sigma_y
+\cos{\chi} \sigma_z)\right)^2,
\end{equation}
where $a$ is the radius of the ring.
The eigenfunctions $\Psi_{n, \pm}$ and eigenvalues $E_{n, \pm}$ of
Hamiltonian (2) in a closed ring are obtained as \cite{b9}
\begin{eqnarray}
\Psi_{n, \pm}& =& \frac{1}{\sqrt{2 \pi}} e^{in \phi}
                \left( \begin{array}{c}
                \cos{\frac{\beta_{\pm}}{2}} \\
                \pm e^{i \phi}\sin{\frac{\beta_{\pm}}{2}}
                \end{array} \right) ~, \nonumber \\
E_{n,\pm}&=&\frac{\hbar^2}{2ma^2} \left(n -
           \frac{\Phi^{\pm}_{\rm AC}}{2 \pi}\right)^2 ~,
\\ \mbox{and}\hspace*{1.5cm}
\Phi^{\pm}_{\rm AC}& =& - \pi(1 - \lambda_{\pm})~,\nonumber
\end{eqnarray}
where $\lambda_{\pm} \equiv \pm \sqrt{\omega_1^2 + ( \omega_3 + 1)^2}$ 
are eigenvalues of $\omega_1 \sigma^1 + (\omega_3 + 1) \sigma^3$, and 
the angle $\beta_{\pm}$ are defined by 
$\tan{\beta_{+}} \equiv \omega_1 / (\omega_3 +1)$, and 
$\beta_{-} = \pi - \beta_{+}$. Here $\omega_1$ and $\omega_3$ are 
denoted by $\omega_1 \equiv \frac{\mu E a}{\hbar c} \sin \chi$ and
$\omega_3 \equiv \frac{\mu E a}{\hbar c} \cos \chi$ and 
$\mu = e \hbar / 2m_e c$ is the Bohr magneton.
The evolution of a spin state in the presence of the electric field is
determined by the following parallel transporter \cite{b9}.
\begin{equation}
 \Omega(\phi)
 = P\exp \Bigg[ i \frac{\mu E a}{2 \hbar c}\int_0^{\phi}
  (\sin{\chi} \cos{\phi'} \sigma^1 + \sin{\chi} \sin{\phi'} \sigma^2
  + \cos{\chi} \sigma^3) d \phi' \Bigg]  ,
\end{equation}
where $P$ is the path ordering operator.
It relates the wave function $\Psi(\phi)$ to $\Psi(0)$.
In general, the spin state that has been parallel transported around 
the ring does not return to the initial spin state. 
However, for the special initial spin state, the spin state after a 
parallel transport around the ring returns to the initial state except 
the phase factor as 
$\Psi (2 \pi) = \exp [i \Phi_{\rm AC}^{(\pm)} ] \Psi(0)$.  
This spin state is the eigenstate of
$\omega_1 \sigma^1 + (\omega_3 + 1) \sigma^3$ \cite{b9}. 
Then the spin state at $\phi$ is obtained as

\begin{equation}
\Psi^{(\pm)}(\phi) = e^{i(1- \lambda_{\pm}) \phi /2}
                     \left( \begin{array}{c}
                \cos{\frac{\beta_{\pm}}{2}} \\
                \pm e^{i \phi} \sin{\frac{\beta_{\pm}}{2}}
                \end{array} \right) ~.
\end{equation}
After a cyclic evolution this spin state returns to the initial
state apart from the AC phase.

To derive an expression for the persistent current and the transmission
coefficient, we apply the one-dimensional quantum
waveguide theory developed in Ref. \cite{b15}.
We use the local coordinate
system for each circuit such that the $x$
coordinate is taken along the electron current flow.
The origin of each  local coordinate
is taken at each junction.
At each junction   charge density and current are conserved,
and electron spins  are matched.
We assume that an electron spin is not changed while
electron passes a junction and neglect the spin-flip process
as in Ref. \cite{b14}.
Since the two reservoirs are mutually phase incoherent, we have to
solve the problem separately for the electrons emitted from the left 
and the right reservoirs.
First we consider the case wherein electrons are emitted from the left
reservoirs. The reservoirs emit electron carriers with the Fermi 
distribution $f(E) = (\exp[(E - \mu_1)/k_B T]+1)^{-1}$. 
This results in a current flowing from the left to the right.

The textured electric field can be made by putting
the extra charge in the center of the ring together with
a circular  gate along the ring.
Then except for the point I (where we have introduced a 
$\delta$-function potential), in the input and output leads, 
there is no normal electric field,
and the Hamiltonian (1) becomes that for the free particle
\begin{equation}
H = \frac{1}{2m_e}p_x^2 ,
\end{equation}
since $p_y =0$ and $E_z=0$ in the leads.
Thus the incident and reflected spin states acquire only the
phases of $i k x$ and $- i k x$,
respectively.
Since it is always possible to use the eigenstates of the ring
at $\phi=0$ ($J_1$) as the basis of general incident spin states, and 
there is no spin flip process, we can treat the eigenstates of the 
ring at $\phi=0$ as the incident spin states separately.
This spin state changes its direction during the movement
along the ring as described in Eq. (5).

When an electron is transported from the input junction in
the clockwise direction along the upper loop, it picks up a phase
$\gamma ={{1 / 2} \Phi_{\rm AC}^\pm}$ at the output junction.
And when the electron is transported in the counter clockwise direction
along the lower loop, the electron
acquires the phase $\delta= {-{1 / 2} \Phi_{\rm AC}^\pm}$.
Thus the total phase around the loop becomes
${ (\gamma -\delta)= \Phi_{\rm AC}^\pm}$.
Since the effect of the electric field on the above spin state brings
the phase shift of wave function, the energy of the electron in the loop
$
E=\hbar^2[k_1^\pm-{\Phi^{\pm}_{\rm AC}}/{2\pi r}]^2/2m
$
should be equal to the energy of the injected electron 
${\hbar^2k^2}/{2m}$.  
Thus we take the wave vector
$k^{\pm}_1=k+{\Phi^{\pm}_{\rm AC}}/{2\pi r}$ for the
electron  moving along the clockwise direction,
and $k^{\pm}_2=k-\frac{\Phi^{\pm}_{\rm AC}}{2\pi r}$  for the electron
moving in the opposite direction.
Let the spin state
$\left( \cos{{\beta_{\pm}}/{2}}~,~
                \pm e^{i \phi} \sin{{\beta_{\pm}}/{2}}  \right)^t $
be $\cal{X}_{\phi}^{\pm}$, where $t$ means the transpose of the vector.
Then the wave functions in the circuits can be written as
\begin{equation}
\left.
\begin{array}{lcl}
\Psi^{\pm}_1&=&({e}^{ikx}+a^{\pm}{e}^{-ikx})
{\cal{X}}_{0}^{\pm},\\
\Psi^{\pm}_2&=&(c^{\pm}_1{e}^{ik^{\pm}_1 x}+
c^{\pm}_2{e}^{-ik^{\pm}_2 x}) \cal{X}_{\phi}^{\pm}, \\
\Psi^{\pm}_3&=&(d^{\pm}_1{e}^{ik^{\pm}_2 x}+
d^{\pm}_2{e}^{-ik^{\pm}_1 x}) \cal{X}_{\phi'}^{\pm}, \\
\Psi^{\pm}_4&=&(f^{\pm}_1{e}^{ik^{\pm}_2 x}+
f^{\pm}_2{e}^{-ik^{\pm}_1 x}) \cal{X}_{\pi}^{\pm}, \\
\Psi^{\pm}_5&=&g^{\pm}{e}^{ik x} \cal{X}_{\pi}^{\pm},
\end{array}\right.
\end{equation}
where the wavefunctions $\Psi_{1-5}$ are for the following regions, 
input lead, $J_1-J_2$ upper arm, $J_1- J_2$ lower arm, $J_2 -I$ 
and output lead, respectively.  
We use the Griffith boundary conditions 
\cite{Griff,Sing1,Jay5} at the junctions.
We have obtained analytical expressions for the persistent currents.
However, here we present our results graphically since the analytical
expression is too lengthy.

In the open system, the transport current is symmetric with respect
to the AC flux. Hence the persistent current is defined as the 
antisymmetric part of the ring current with respect to the AC flux.
As is well known, the Hamiltonian considered has the time-reversal
symmetry. Because of this time reversal symmetry the persistent 
charge currents for the unpolarized incident electrons always vanish. 
In the presence of the net spin polarization the AC effect leads to
charge currents proportional to $n_\uparrow - n_\downarrow$, where
$n_\uparrow$ and $n_\downarrow$ are the number of spin up electrons 
and down electrons, respectively. 
In Fig. 2 we have plotted the persistent charge currents for the 
incident spin-up eigenstate of the ring at $\phi=0$ in the
dimensionless unit $J/k$ as a function of the normalized field
strength $\eta$ for tilt angles $\chi=0$(A), $\pi/2$(B), $3 \pi/4$(C),
and $\pi$(D), the dimensionless 
momentum $kL=7$, and the impurity strength $VL=10$. 
In Fig. 2 the solid and dashed curves represent the magnitudes 
of the persistent charge currents flowing in
the loop ${J_C^+}^L /k$, and ${J_C^+}^R /k$, respectively for the 
spin-up eigenstate of the ring. 
Where the subscript L and R represent when
the dc current flows in the left and right directions,
respectively. One can readily notice from Fig. 2 the difference
between the values of the persistent charge current for
the electron emitted from the left reservoir (solid line) and that
for the electron emitted from the right reservoir (dashed line). 
This shows clearly that the
persistent charge currents in a metal loop connected to two
reservoirs depend on the direction of direct current flow from one
reservoir to the other. For $V=0$ we can recover the symmetric
case, so there is no directional dependence on the persistent
charge currents. We can also see the anomalous behaviors in the
persistent charge current when the normalized electric field $\eta$
is small and $\chi \ne 0$, $\pi$. This anomalous behavior comes
from the nonadiabatic AC phase.
The persistent charge currents
for the spin-down eigenstate of the ring is exactly opposite to
the spin-up persistent charge currents because of the time
reversal symmetry.

The persistent spin current $J_S^a$ is defined as the antisymmetric part
of $<\Psi_r| ({\bf p} - \mu r /c \cdot {\bf \sigma }
   \times {\bf E})_{\phi} \sigma^a / \hbar |\Psi_r>$. Where
   $a=1$, $2$ and $3$ is the spin indices and $|\Psi_r>$ is the
   state on the ring.
It has the additional contribution from $\sigma^a$ operator to the
persistent charge current. Because of the cylindrical symmetry, 
the persistent spin current with the x and y direction vanishes. 
From $<\Psi_{n,\alpha}|\sigma^3|
\Psi_{n, \alpha}>= \cos{\beta_{\alpha}}$, and $\cos_{\beta_-}
=-\cos{\beta_+}$, the persistent spin current $<J_S^3>$ of 
spin-down eigenstate
of the ring is the same as that of spin-up eigenstate.
This implies that the persistent spin current would be independent
of spin polarization. It should be noticed that the magnetic field
necessary for spin polarization is not required to observe the
persistent spin current, different from the persistent charge
current. 
Fig. 3 shows the persistent spin current as a function of normalized
electric field strength $\eta$ for the same values as that of
the persistent charge current. In the case of $\chi \ne 0$,
$\pi$, the anomalous behavior is more definite than that of 
the persistent charge current, because of the additional
contribution from the modulation $\cos \beta_+$.

We give a simple picture to understand the anomalous 
behavior of persistent spin current intuitively.
We consider the spin up eigenstate only
in the following since SO interaction term is time-reversal invariant.
It is also the eigenstate of $\omega_1 \sigma^1 +
(\omega_3 + 1) \sigma^3$ with spin up. 
In a ring the system has a cylindrical symmetry, so the spin 
direction at $\phi$ has the polar angle $\beta_{+}$ and 
the azimuthal angle $\phi$.
It means that the spin precesses about $\hat{z}$ direction with an angle
$\beta_{+}$ during the cyclic evolution.
From the similarity of the mathematical structure of the AC effect
with the AB effect we can rewrite
the spin-orbit coupling term as the effective spin dependent gauge
field $\frac{e}{c}{\bf A}_{\rm eff}$, with ${\bf A}_{\rm eff}=
\frac{\mu}{\hbar e} ({\bf S} \times {\bf E})$.
Where ${\bf S}$ is the spin operator.
In the semi-classical approach a spin is a  three-dimensional
vector with a certain direction.
The $\frac{\mu}{\hbar e} ({\bf S} \times {\bf E})$ is calculated as
$\frac{\mu E}{2 e} \cos(\beta_{+} - \chi) \hat \phi$. 
This is constant during the motion as far as the field strength 
$E$ and the tilt angle $\chi$ is fixed.
Hence this $\frac{\mu}{\hbar e} ({\bf S} \times {\bf E})$ is described 
as ${\bf A}_{\rm eff} = \frac{\Phi}{2 \pi a} \hat{\phi}$.
Where $\Phi = (\nabla \times {\bf A}_{\rm eff}) \cdot {\bf F} $
is the magnetic flux through the ring section area $F(= \pi a^2)$.
The phase acquired from this effective AB situation - we call this
$\Phi^{\rm eff}_{\rm AB}$ - is

\begin{equation}
\Phi^{\rm eff}_{\rm AB} = 
	   \frac{\pi e a^2 E}{2 m_e c^2} \cos(\beta_{+} - \chi).
\end{equation}

This effective spin dependent AB phase acquired by a charge $e$ around 
a flux $\Phi=(\nabla \times {\bf A}_{\rm eff}) \cdot {\bf F}$
turns out to be the same as the dynamical phase acquired by a spin
due to the SO interaction \cite{b13,b9}.
In the AB situation
a charge does not precess, but in the AC situation the spin precesses
during the cyclic evolution,  bringing an additional effect.
The difference between the effective AB phase for a charge  and
the AC phase  for a spin becomes the AA phase, which comes from 
the extra spin degrees of freedom.

The AA phase is  associated with the spin precession.
To get this phase we parametrize the path  of the spin by the azimuthal
angle $\phi$. The spin state
$| {\bf S} \cdot \hat{n} ; + >$ satisfies
\begin{equation}
{\bf S} \cdot \hat{n} ( \phi ) | {\bf S} \cdot \hat{n} ; + >
 ~= ~ \frac{\hbar}{2} | {\bf S} \cdot \hat{n} ; + >~,
\end{equation}
where $\hat{n} ( \phi )$ is the unit vector with polar angle 
$\beta_+$ and azimuthal angle $\phi$. 
And the spin state $| {\bf S} \cdot \hat{n} ; + >$
becomes ${\cal{X}}_{\phi}^{+}$.
Since this spin ${\bf S}$ remains parallel to $\hat{n}(\phi)$ during the
rotation, formally this is identical to the problem considered by Berry
for a spin ${\bf S}$ in an adiabatically changing magnetic field 
${\bf B}(t)$.  
\begin{equation}
g {\bf S} \cdot {\bf B} (t) | {\bf B} (t), ~ m_s > ~
= ~ E | {\bf B} (t), ~ m_s >
\end{equation}
where $g$ is related to the gyromagnetic ratio and $m_s$ is the 
component of the spin along the direction of ${\bf B} (t)$.
Berry showed that $ \gamma (C) ~= ~ -m_s \Omega(C)~$, 
where $\gamma(C)$ is Berry's phase and $\Omega (C)$ is the solid 
angle subtended by the curve $C$ with respect to the origin 
${\bf B} ~= ~0$.
In our case, the phase accumulated is
\begin{equation}
\gamma (C) ~= ~ - \frac{1}{2} \Omega(C)~,
\end{equation}
where $\Omega (C)$ is the solid angle subtended by the loop $C$ with 
respect to ${\bf n} ~= ~ 0$.
In this case $C$ is a circle and
$\Omega(C) ~= ~ 2 \pi (1-\cos{\beta_+})$.
This geometric phase $\gamma (C)$  is the AA phase, and thus the
AA phase becomes  $- \pi (1 - \cos{\beta_{+}})$.

From the above intuitive picture, the anomalous behavior is
understandable by the precession of the spin. 
The AA phase is determined by the solid angle of spin precession
and the dynamical phase is the effective AB phase induced by the
spin dependent AB flux.
Let us first consider the adiabatic approximation of the spin evolution.
The condition for the adiabatic limit is $\eta \gg 1$.
In this case the spin state is an eigenstate of
the parallel transporter. The dynamical phase of
the adiabatic solution is given by
$\Phi_{\rm dyn}^{\pm} \approx \pm \sqrt{\omega_1^2 + \omega_3^2}$
from Eq. (8). Also the adiabatic approximation of the AA phase is the
Berry phase, $\Phi_{\rm Berry}^{\pm} = - \pi (1 \mp \cos{\chi})$.
These phases are equal in Ref. \cite{b14} for proper parameter 
transformation.  In this limit, the spin precession angle $\beta_+$
becomes the fixed tilt angle $\chi$.
The AA phase gives a constant shift to AC phase.
That is, the effective spin dependent $\Phi_{AB}^{eff}$ determines
the periodic behaviors.
For $\chi=0$, the AC phase consists of the dynamical phase only 
and persistent spin current oscillates periodically. 
It is clear that the anomalous behaviors of the persistent spin
currents come from the change of the spin precession angle with
varying electric field.
For a fixed tilt angle the precession angle $\beta_+$ of spin-up
depends only on the field strength $\eta$ as
$$
\cos{\beta_+}= \frac{\eta\cos{\chi} + 1}{\sqrt{\eta^2 \sin{\chi}^2
+(\eta \cos{\chi} + 1)^2}}.
$$
For $\eta>0$, this can be negative for $\chi> \pi/2$.
We can see this change of sign of the persistent spin current 
in Fig 3 (C) and (D) in comparison with the persistent charge current.

In Fig. 4 we have plotted the persistent spin currents 
${J_S^+}^{R} /k$ and ${J_S^+}^{L} /k$ as a function of dimensionless 
impurity potential $VL$, 
for a fixed value of $kL=7$, for $\eta=6$ and for tilt angle $\pi/2$. 
In this case the modulation $\cos \beta_+$ has a fixed value $0.16$. 
The magnitude ${J_S^+}^R /k$ decreases monotonically to zero as 
$VL\rightarrow \infty$. This is due to the fact that in this limit 
electrons emitted by the right reservoir do not enter the loop and 
cannot contribute to the persistent spin currents. 
The absolute magnitude of ${J_C^+}^L /k$ saturates to a value in the 
same limit. 
This corresponds to a situation where the loop is connected to 
a single reservoir $\mu_L$, where the connection is truncated at 
the point $I$ (the impurity state). 
In Fig. 5 we have plotted the dimensionless conductance $T$ ($|g^+|^2$)
as a function of $kL$, for $\eta=6$, for $VL=10$ and for
$\chi=\pi/2$. The electrical conductance exhibits a peak for
certain values of $kL$. These peaks occur due to the resonance of
the incident electron energy coincides with one of the 
eigenenergies of the ring or with one of the bound state energies of 
the stub $J_2-I$. But the peaks do not appear at the exactly same
energy as the eigenenergies since the multiple scatterings at 
junctions shift the energy levels.
And we can see the effect of the bound state of the stub $J_2-I$ will
decrease as $kL$ becomes much higher than $VL$.

In conclusion, we have shown that the magnitude of the persistent spin 
current induced by the Aharonov-Casher phase
in a normal metal loop connected to two reservoirs depends on the 
direction of the direct current flow, which should be an 
experimentally verifiable feature. 
In the presence of the AC flux, the Hamiltonian has the time reversal
symmetry, so the persistent charge current always vanishes for 
unpolarized incident electrons. But the persistent spin current still
exists. That is, if the spin-up electrons of the ring circles 
counter-clockwise, then the spin-down electrons evolves clockwise
and vise versa.
Since the conductance of the entire network does not depend on 
the direction of the direct current flow,
there is no simple scale relationship between the persistent spin
currents and the conductance of the entire network.
The anomalous behaviors appear when the spin precession angle changes
as a function of the field strength.
The periodic behavior in the adiabatic limit can be understood
as the effect of spin dependent AB flux. 
In this case the spin precession angle does not change so that the AA 
phase is constant.  
The difference between the magnitude of the persistent spin currents 
(on the direction of the current flow) can be made significant by 
adjusting the impurity potential. 
This can be achieved experimentally by having a gate in one of the 
leads connected to the reservoirs and by appropriately varying   
the gate voltage.
Such an experiment can also be useful for separating the persistent spin
currents from other parasitical currents (or signals) associated with
measurements.
When there are time reversal symmetry breaking terms in the Hamiltonian,
we expect that the persistent charge current will also appear
even for unpolarized incident electrons.
The natural terms are the Zeeman coupling and the AB flux of 
the localized magnetic field.
The directional dependence of spin and charge currents 
in the presence of the Zeeman coupling, AB flux and 
AC flux is currently under study by us.

A. M. Jayannavar acknowledges for the hospitality at POSTECH.
One of us (Choi) acknowledges S. Y. Cho for helpful discussions.
This work was supported in part by Korean Research Foundation,
POSTECH BSRI special fund, and KOSEF.

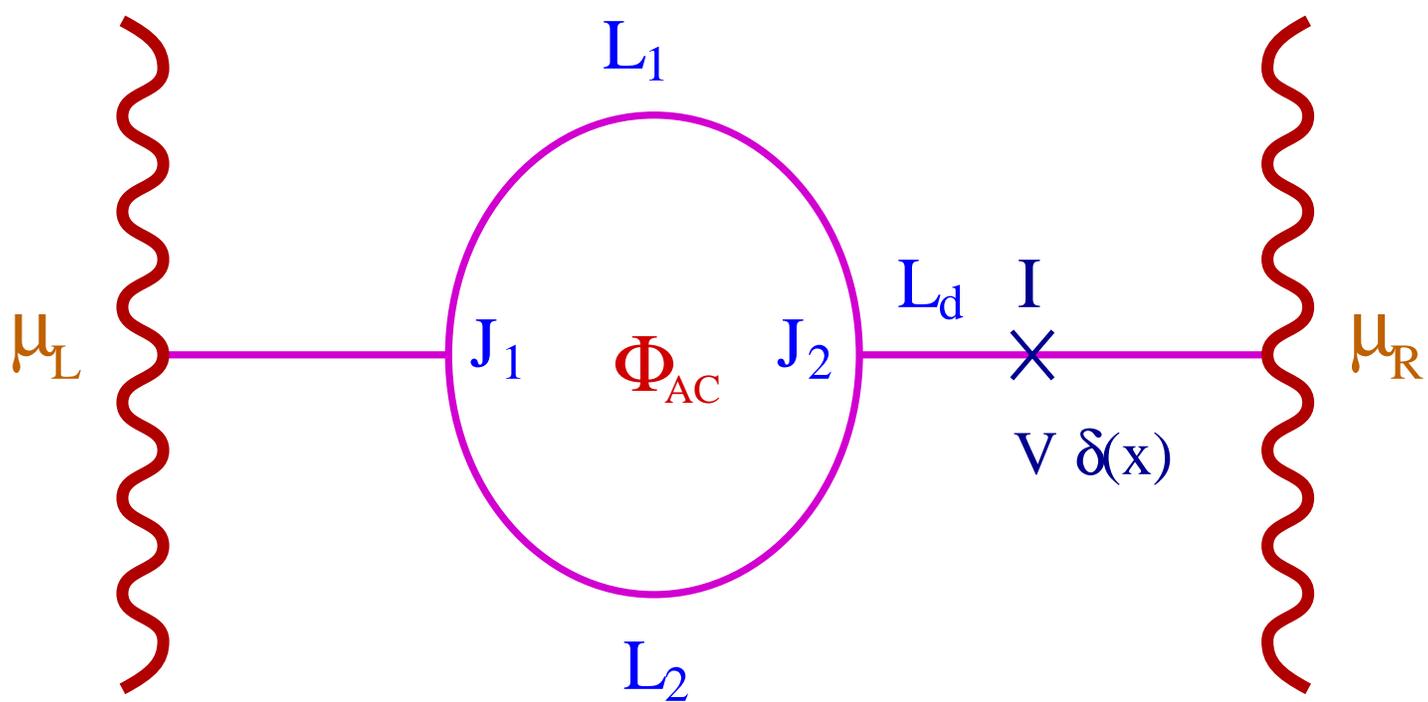
\begin{figure}
\caption{An open metallic loop connected to two electron reservoirs.
There exist a cylindrically symmetric electric field which gives
the AC flux.}
\label{fig1}\end{figure}

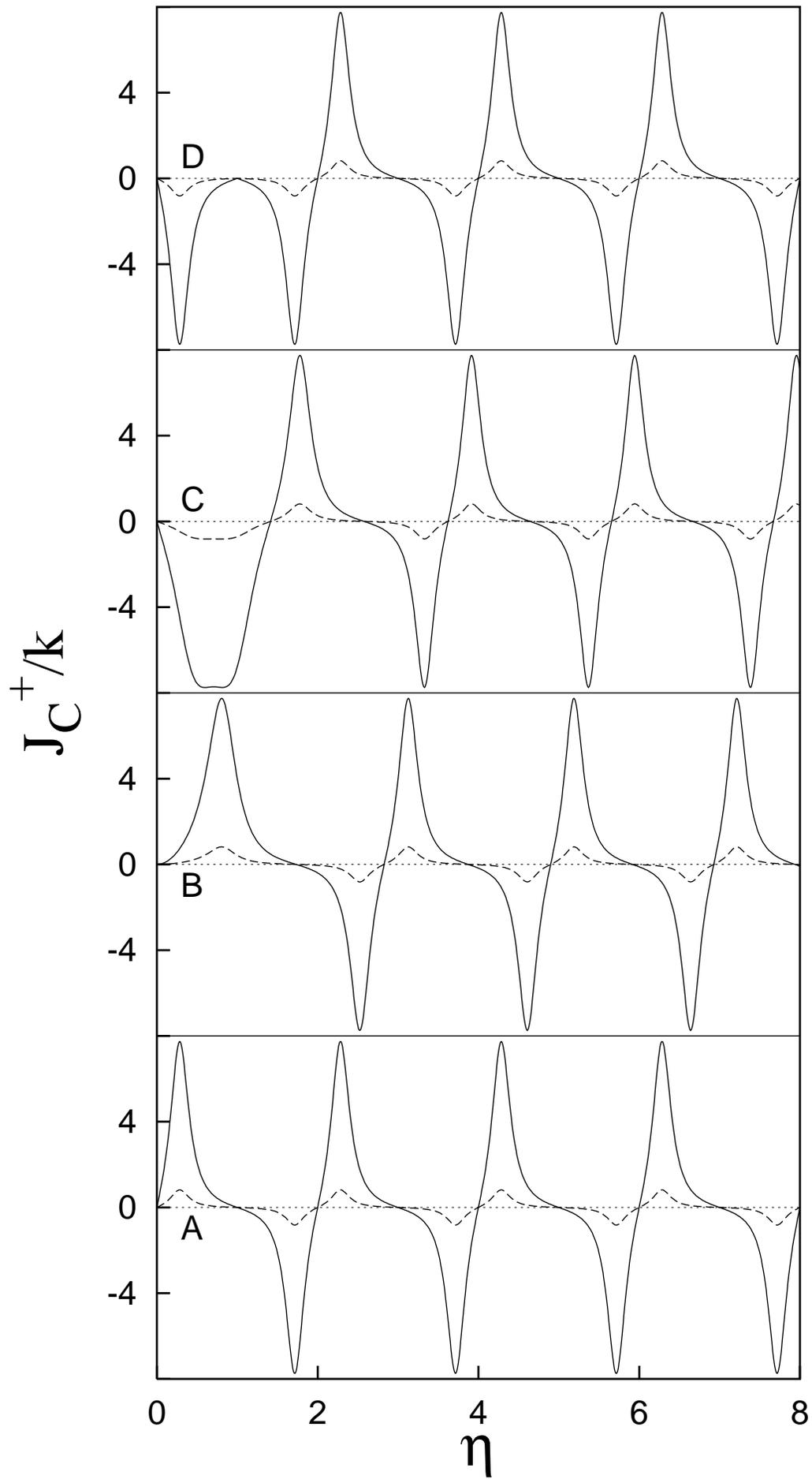
\begin{figure}
\caption{The persistent charge current as a function of the normalized
electric field $\eta$ for a fixed value of $kL=7$, $VL=10$, tilt angles
(A) $\chi=0$, (B) $\pi/2$, (C) $3 \pi/4$, and (D) $\pi$.
The solid line represents persistent charge current for ${J_C^+}^L /k$
and the dashed curve represents ${J_C^+}^R /k$.}
\label{fig2}\end{figure}

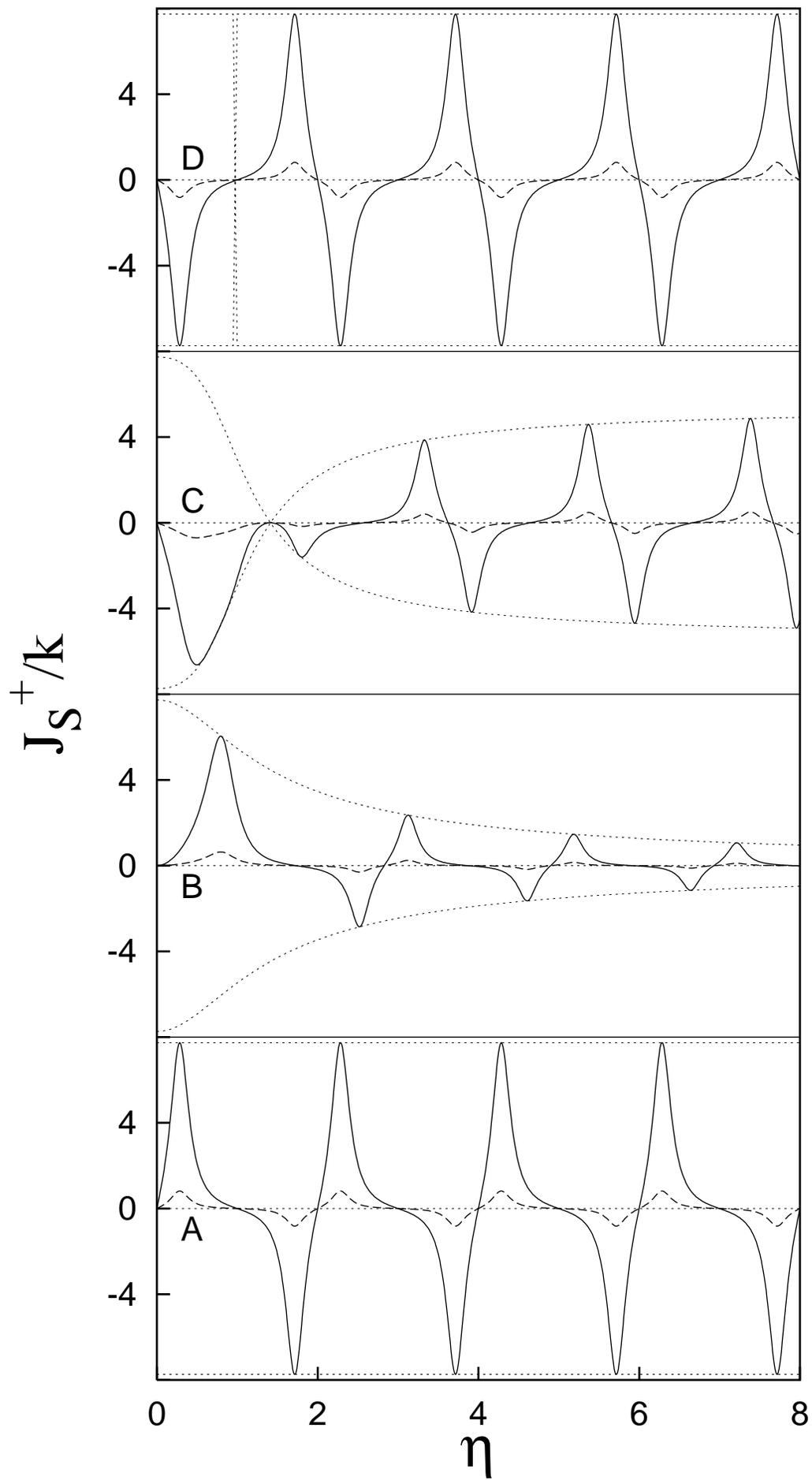
\begin{figure}
\caption{The persistent spin currents vs $\eta$ for same values 
in Fig. 2.
The solid line represents ${J_S^+}^L /k$ and dashed curve 
represents ${J_S^+}^R /k$. 
And the dotted line represents the modulation function as an envelope.}
\label{fig3}\end{figure}

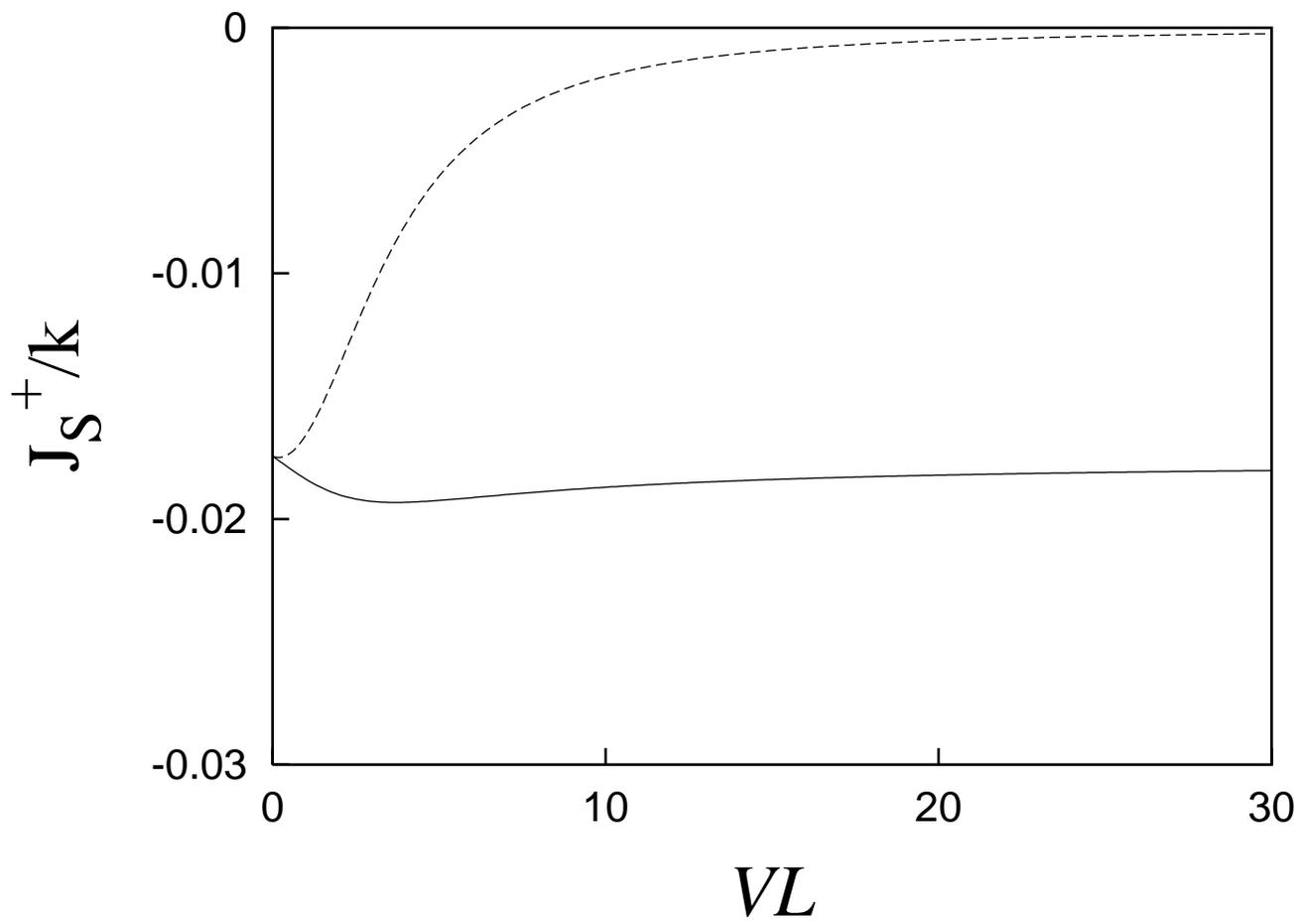
\begin{figure}
\caption{The persistent spin currents vs the strength of impurity 
potential for fixed values of $kL=7$, $\eta=6$ and $\chi = \pi/2$. 
The solid line represents ${J_S^+}^L /k$ and dashed curve represents 
${J_S^+}^R /k$.}
\label{fig4}\end{figure}

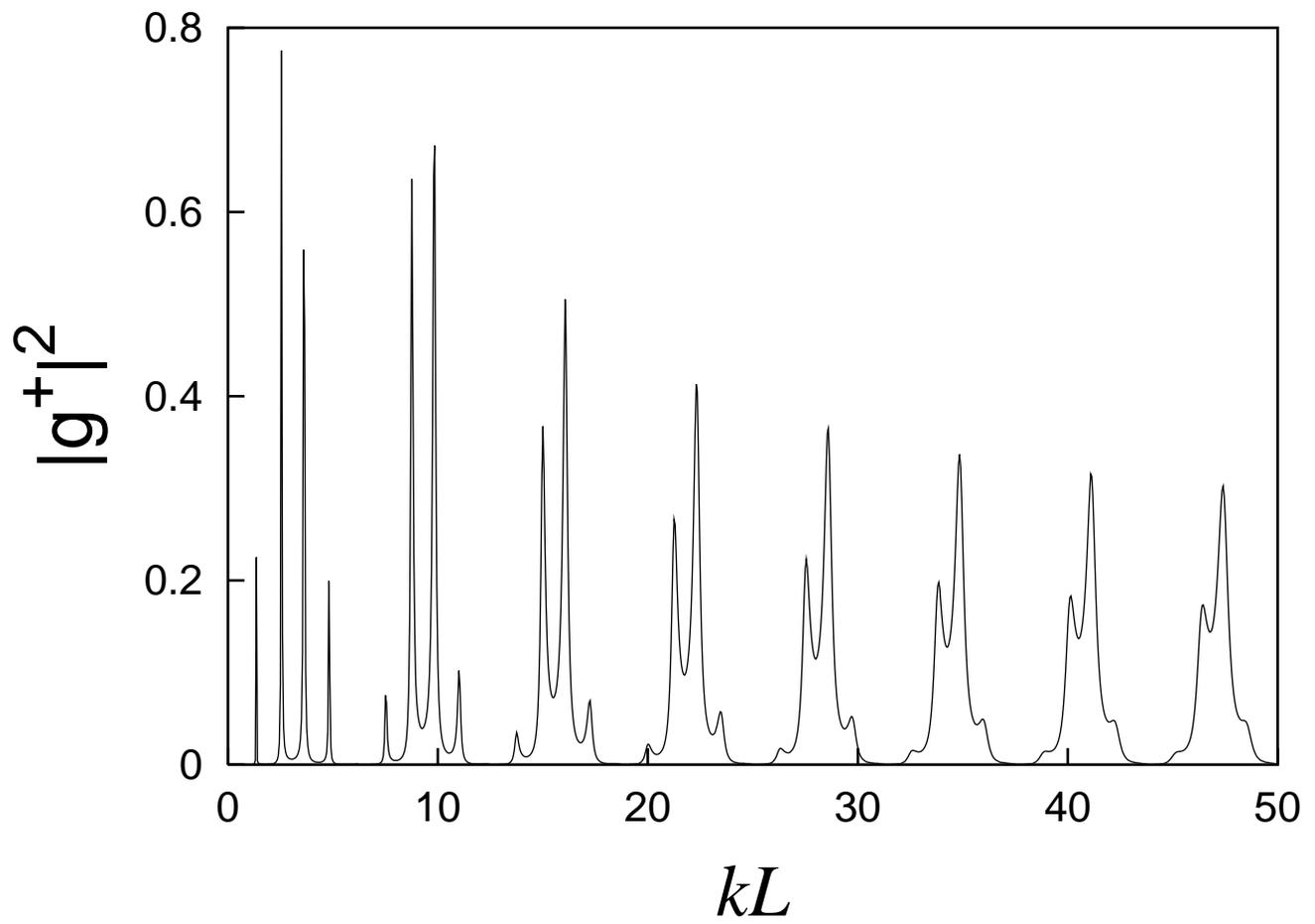
\begin{figure}
\caption{Conductance oscillations vs $kL$ for $\eta=6$, $VL=10$
and $\chi\ \pi/2$.}
\label{fig5}\end{figure}

\end{document}